\documentstyle[11pt]{article}
\newcommand{\be}{\begin{equation}}
\newcommand{\ee}{\end{equation}}
\newcommand{\ba}{\begin{eqnarray}}
\newcommand{\ea}{\end{eqnarray}}
\newcommand{\demu}{\partial_{\mu}}
\newcommand{\denu}{\partial_{\nu}}
\newcommand{\bea}{\begin{array}{l}}
\newcommand{\eea}{\end{array}}
\newcommand{\ds}{\displaystyle}
\newcommand{\ldl}{\Lambda \partial_{\Lambda}}
\begin{document}

\title{{\bf \huge{Renormalization group flow for Yang-Mills\\ fields interacting 
with matter}}}
\author{Stefano Arnone\thanks{e-mail: Stefano.Arnone@roma1.infn.it}\hspace{.5em}
and Annamaria Panza\thanks{e-mail: Panza@vaxsa.csied.unisa.it}\\
Department of Physics, University of Rome ``La Sapienza''\\
Piazzale Aldo Moro, 5 - Roma, Italy}
\maketitle
\begin{abstract}
We show an application of the Wilson Renormalization Group (RG) method to a
$SU(2)$ gauge field theory in interaction with a massive fermion doublet. By 
choosing suitable boundary conditions to the RG equation, \emph{i.e.} by requiring
the relevant monomials not present in the classical action to satisfy the Slavnov-
Taylor identities once the cutoffs are removed, we succeed in implementing the local gauge symmetry. In this
way the so called fine-tuning problem, due to the assignation of boundary 
conditions in terms of the bare parameters, is avoided. In this framework, loop
expansion is equivalent to the iterative solution of the RG equation; we perform
one loop calculations in order to determine whether and, if so, how much the 
fermionic matter modifies the asymptotic form of the couplings. Then we compute
the $\beta$-function and we check gluon transversality. Finally, a proof of
perturbative renormalizability is shown.
\end{abstract}
\vspace{3 truecm}

\newpage
\section{Introduction}
For a long time U.V. divergences appearing in gauge field theories have been 
treated by those regularizations which were able to preserve the local gauge 
symmetry; in this way, indeed, it has been possible to take into account only
the counter terms the symmetry itself allowed.\\
This is why dimensional regularization \cite{RA} is the most used one in 
dealing with such a theory.\\
However, dimensional regularization is bounded to perturbative regimes and,
moreover, it cannot be easily extended to chiral gauge theories, since one
should define Dirac'~s matrix $\gamma_{5}$ in complex space-time dimensions.\\
Therefore, it would be useful to find a new regularization scheme avoiding
these problems: the Wilson renormalization group (RG) formulation \cite{Wi} has
just these features since it works in four dimensions and it is, in principle,
non-perturbative.
Nevertheless, gauge symmetry breaking, due to the cutoff introduction, is the
price to be payed.\\
Polchinski, in 1984, \cite{Po} proposed an original way of computing Green
functions by a suitable RG equation.\\
A few years later, Becchi \cite{Be} analysed the breaking induced by the cutoff
in a $SU(2)$ gauge field theory and found a way of compensating it by a
suitable choice of non gauge-invariant counter terms (fine-tuning equation).\\
Recently Bonini et al. \cite{SU2} succeeded in avoiding  the fine-tuning problem by
choosing in a proper way boundary conditions to the RG equation.\\
In this paper we extend RG formulation to a $SU(2)$ gauge field theory
interacting with a fermionic doublet by a minimal coupling.\\
The first step is the analysis of the classical action ($S_{cl}$) invariance
under BRS transformations \cite{BRS}. 
This invariance forces Green functions to satisfy the Slavnov-Taylor (ST)
identities, the analogous of Ward'~s in the non-abelian case.
These identities can be expressed in a compact form by a functional identity
generating them by successive functional derivations.\\
The second step is the introduction of a regularization scheme consisting in
assuming the free propagators in momentum representation to vanish for $p^{2}
> \Lambda_{0}^{2}$ and $p^{2} < \Lambda^{2}$ so that $\Lambda$ and
$\Lambda_{0}$ represent respectively the infrared and ultraviolet cutoffs. 
This procedure is equivalent to Wilson'~s since the Wilsonian effective action
generates propagators having the same features.
Obviously, the cutoff introduction breaks the gauge symmetry; as a consequence,
a suitable set of non gauge-invariant counter terms has to be added and,
moreover, the
cutoff effective action, $\Gamma \{ \underline{\varphi}, \Lambda \}$, does not satisfy the 
ST identities at the generic scale $\Lambda$.\\
The successive step is the derivation of the RG equation together with boundary
conditions, which can be fixed in different ways. One can assign the bare
parameters, \emph{i.e.} the couplings at the U.V. scale, giving rise to the 
fine-tuning problem, or, in order to avoid it, one can fix different boundary
conditions according to the different nature of the couplings.\\
The irrelevant couplings, with negative mass dimensions, are assumed to vanish
at the U.V. scale, while the relevant ones, with non-negative mass dimensions,
are fixed at the physical point $\Lambda = 0$, $\Lambda_{0} \rightarrow \infty$
to be the physical masses and the coupling and wave function constants.
Since the interaction generates also terms with no classical analogous, the
relative conditions are given by requiring the relevant part of the effective
action to satisfy the ST identities at the physical point; this is the first step
towards gauge symmetry recovery at the quantum level.\\
The RG equation obtained in this way is exact; once boundary conditions are
given, it allows to compute all the vertices of the theory (and, so, all the 
physical observables) either by using numerical techniques or by solving it
perturbatively.\\
In the last framework we compute, at one loop, all the couplings involved in
the vertex functions; starting from them, we derive the one loop
$\beta$-function, related to the asymptotic freedom, and we check the gluon
transversality as a consequence of one of the ST identities in the physical
limit.\\
This paper is organized in the following way: in section $2$ we analyse BRS
symmetry and its consequences over Green functions; then we derive the RG
equation and we fix suitable boundary conditions. In section $3$ we discuss in
detail the fermionic sector, only quoting the other results. In section $4$ we
give a schematic proof of perturbative renormalizability.

\section{Renormalization group flow}
\subsection{Classical action, BRS symmetry and ST identities}
The classical euclidean action in the Feynman gauge is 
\be \label{scl}
S_{cl}(A_{\mu}, \bar{c}, c, \bar{\Psi}, \Psi) = \int dx \, \left[ i \bar{\Psi}\,
(i D \hspace{-0.6 em}/ - m)\, \Psi + \frac{1}{4} F_{\mu
\nu} \cdot F_{\mu \nu} + \frac{1}{2}(\demu A_{\mu})^{2} + 
\bar{c} \cdot \demu {\cal D}_{\mu} c \right]
\ee
where
\be
\bea
{\ds D_{\mu} \doteq \demu - i g A_{\mu} }\nonumber\\
{\ds F_{\mu \nu} \cdot F_{\mu \nu} \doteq F_{\mu \nu}^{a} F_{\mu \nu}^{a}
 }\nonumber\\
{\ds F_{\mu \nu}^{a} \doteq \demu A_{\nu}^{a} - \denu A_{\mu}^{a} + g
(A_{\mu} \wedge A_{\nu}^{c})^{a} } \nonumber\\ 
{\ds {\cal D}_{\mu} c \doteq \frac{1}{g} \demu c + A_{\mu} \wedge c }\nonumber
\eea
\ee
the last two terms in (\ref{scl}) representing respectively the gauge-fixing
and the Faddev-Popov determinant parametrization.\\
This action is invariant under the non-linear, nihilpotent BRS transformations
\be \label{brs}
\bea
{\ds A_{\mu} = A'_{\mu} +\tilde{\eta} {\cal D}_{\mu} c' }\nonumber\\
{\ds \bar{c} = \bar{c'} -\tilde{\eta} \demu A'_{\mu} }\nonumber\\
{\ds c = c' -\frac{\tilde{\eta}}{2} c' \wedge c' }\nonumber\\
{\ds \Psi = \Psi' + i \tilde{\eta} c' \Psi'}\nonumber\\
{\ds \bar{\Psi} = \bar{\Psi'} + i \tilde{\eta} \bar{\Psi'} c'}
\eea
\ee
$\tilde{\eta}$ being a Grassmann parameter.\\
Owing to this invariance, vertex functions have to satisfy the ST identities, whose
functional formulation can be simplified by adding to $S_{cl}$ other four
terms, obtained as scalar products between field variations and their coupled
sources\footnote{(\ref{sbrs}) is still invariant under (\ref{brs})
because of the BRS operator nihilpotency.}.
\be \label{sbrs}
S_{BRS} \doteq S_{cl} - \int dx \, \left( k_{\mu} \cdot {\cal D}_{\mu} c +
\frac{1}{2} L \cdot c \wedge c + i \bar{\lambda} c \Psi + i \bar{\Psi} c
\lambda \right) 
\ee
The generating functional of Green functions is
\be \label{fungen}
Z\{ \underline{j} \} = \exp W\{ \underline{j} \} = \int {\cal D} \underline{\varphi} \exp \left[ -S_{BRS}
[\underline{\varphi}] + \int dx j_{A}(x) \varphi_{A}(x) \right]
\ee
where
\be
\underline{\varphi} \doteq (A_{\mu}, c, \bar{c}, \Psi, \bar{\Psi})  \nonumber
\ee
\be
\underline{j} \doteq (j_{\mu}, \bar{X}, -X, \bar{\eta}, -\eta) \footnote{This
form of $\underline{j}$ takes into account the exchange properties of fields and
sources.} 
\ee
In order to obtain the ST identities we perform the substitution (\ref{brs}) in 
(\ref{fungen}), noting the invariance of the integration measure and of the
action as well.\\
By a first order $\tilde{\eta}$-expansion\footnote{The expansion can be made up
to the first order because of the Grassmann nature of $\tilde{\eta}$.} we obtain
\be \label{relbrs}
\int dx \, \left[ j_{\mu} \cdot \frac{\delta}{\delta k_{\mu}}
 + \bar{\eta} \cdot \frac{\delta}{\delta L} - \demu \frac{\delta}{\delta
j_{\mu}} \cdot \eta - \bar{X} \frac{\delta}{\delta \bar{\lambda}} +
 X \frac{\delta}{\delta \lambda} \right] Z\{\underline{j}\} = 0
\ee
(\ref{relbrs}) can be expressed in terms of the effective action $\Gamma$,
obtained by taking the functional Legendre transform of $W\{ \underline{j} \}$,
defined in (\ref{fungen}).
\be \label{relbrsgammaprimo}
\int dx\, \left[ -\frac{\delta \Gamma'}{\delta A_{\mu}} \cdot \frac{\delta
\Gamma'}{\delta k_{\mu}} + \frac{\delta \Gamma'}{\delta c} \cdot \frac{\delta 
\Gamma'}{\delta L} - \frac{\delta \Gamma'}{\delta \Psi} \frac{\delta \Gamma'}
{\delta \bar{\lambda}} - \frac{\delta \Gamma'}{\delta \bar{\Psi}} \frac{\delta
\Gamma'}{\delta \lambda} \right] = 0
\ee
where ${\ds \Gamma' \doteq \Gamma + \frac{1}{2} \int dx \, 
(\demu A_{\mu})^{2} }$.\\
The ST identity validity holds a great importance in order to assign the
boundary conditions to the RG flow equation.

\subsection{RG equation}
In order to regularize the theory we will adopt a scheme formally equivalent to
Wilson'~s; it consists in multiplying the quadratic part of the BRS action by a
cutoff function $K_{\Lambda \Lambda_{0}}(p)$\footnote{$K_{\Lambda \Lambda_{0}}
(p)$ regularity is treated in \cite{Morris}.}, defined in momentum space as:
\be
\begin{array}{lll}
K_{\Lambda \Lambda_{0}}(p) = 1 &    &\Lambda^{2} \leq p^{2} \leq
\Lambda_{0}^{2} \nonumber\\
supp K_{\Lambda \Lambda_{0}} \subset A &    &A \supset \{ p\,|\ \Lambda^{2} \leq p^{2} \leq
\Lambda_{0}^{2} \}
\end{array}
\ee
The resulting cutoff action $S^{\Lambda \Lambda_{0}}$ consists of two terms:
$S_{2}^{\Lambda \Lambda_{0}}$, which represents the quadratic part of the
action, and $S_{int}^{\Lambda_{0}}$, which contains all the Lorentz scalar
$SU(2)$ singlet monomials whose dimensions are not higher than four, according
to the power counting.\\
The symmetry breaking, due to the cutoff introduction, makes the interaction
generate even non gauge-invariant monomials, which, of course, have no
classical analogous.
\be 
\bea
{\ds S_{2}^{\Lambda,\Lambda_{0}} = \int dx dy \, K^{-1}(y-x) \left[
-\frac{1}{2} A_{\mu}(y) \cdot \partial^{2}_{x} A_{\mu}(x) +
 i \bar{\Psi}(y)\, (i \partial \hspace{-0.6 em}/ - m)_{x} \Psi(x) + \right.
}\nonumber\\
{\ds + \left.\frac{1}{g} \bar{c}(y) \cdot \partial^{2}_{x}  c(x) \right] \doteq 
\frac{1}{2} \int dx dy \, \varphi_{A}(y) D^{-1}_{AB}(y,x;\Lambda)
\varphi_{B}(x) }
\eea
\ee
\be \label{sint}
\bea
{\ds S_{int}^{\Lambda_{0}} = \int dx \, \bigg[
\frac{1}{2} A_{\mu} \cdot
\Big( \delta_{\mu \nu} ( \sigma_{m_{A}}(\Lambda_{0}) - \sigma_{\alpha}
(\Lambda_{0}) \partial^{2} ) -
\sigma_{A}(\Lambda_{0}) ( \delta_{\mu \nu} \partial^{2} - \demu \denu ) \Big)
A_{\nu} + }\nonumber\\
{\ds + i \bar{\Psi} \, \Big( -\sigma_{m_{\Psi}}(\Lambda_{0}) + \sigma_{\Psi}
(\Lambda_{0})(i \partial
\hspace{-0.6 em}/ -m) \Big) \,\Psi - \sigma_{\omega c}(\Lambda_{0}) \omega_{\mu}
\cdot \demu c + \sigma_{3A}(\Lambda_{0}) \demu A_{\nu} \cdot  }\nonumber\\
{\ds \cdot (A_{\mu} \wedge A_{\nu}) + \frac{\sigma_{4A}(\Lambda_{0})}{4} 
(A_{\mu} \wedge A_{\nu}) \cdot (A_{\mu} \wedge
A_{\nu}) + \sigma_{\omega c A}(\Lambda_{0}) \omega_{\mu} \cdot (c \wedge 
A_{\mu}) + }\nonumber\\
{\ds + i \sigma_{\bar{\Psi} A \Psi}(\Lambda_{0}) \bar{\Psi} A \hspace{-0.6 em}/ 
\Psi -\frac{\sigma'_{Lcc}(\Lambda_{0})}{2}
L \cdot c \wedge c - i \sigma'_{\lambda c \Psi}(\Lambda_{0}) \left[ 
\bar{\lambda} c \Psi +
\bar{\Psi} c \lambda \right] +}\nonumber\\
{\ds + \frac{\sigma'_{4A}(\Lambda_{0})}{8} \Big[ 2(A_{\mu} \cdot A_{\nu})(A_{\mu}
 \cdot A_{\nu}) + (A_{\mu} \cdot
 A_{\mu})(A_{\nu} \cdot A_{\nu}) \Big] \bigg] }
\eea
\ee
By $S^{\Lambda \Lambda_{0}}$ the regularized generating functional $Z\{\underline{j},
\Lambda \}$\footnote{Only the $\Lambda$-dependence will be explicitly written in
$Z$ as well as in the other functionals, since we are always interested in the
limit $\Lambda_{0}\rightarrow \infty$.} is constructed and by noting that all
its $\Lambda$-dependence comes from $K_{\Lambda \Lambda_{0}}(p)$ the RG
equation is easily derived.
\be \label{equazionegruppo}
\ldl \Pi\{\underline{\varphi};\Lambda\} = -\frac{1}{2} \int dk dl \, M_{ED}(l,k)
\bar{\Gamma}_{DE}(k,l)
\ee
where
\be
\bea
{\ds \Pi\{\underline{\varphi};\Lambda\} \doteq - \Gamma\{\underline{\varphi};\Lambda\} - S_{2}^{\Lambda
\Lambda_{0}} }\nonumber\\
{\ds M_{ED}(l,k) \doteq \int dx dy \, \Delta_{EA}(l-y) \left[ \ldl
D^{-1}_{AB}(y,x;\Lambda) \right] \Delta_{BD}(x-k)}\nonumber\\
\eea
\ee
$\Delta_{AB}(x-y)$ being the full propagator.
The auxiliary functional $\bar{\Gamma}_{BF}$ satisfies the integral equation
\be \label{eqdigammabarra}
\bar{\Gamma}_{BF}(y,\xi;\underline{\varphi}) = (-1)^{\delta_{F} + 1} \Gamma^{int}_{BF}
(y,\xi;\underline{\varphi}) +
\int dk dz \, \Delta_{CD}(z-k) \bar{\Gamma}_{DF}(k,\xi;\underline{\varphi}) \Gamma^{int}_
{BC}(y,z;\underline{\varphi}) 
\ee
where $\Gamma^{int}_{BC}(y,z;\underline{\varphi})$ is defined by the relation
\be
\frac{\delta^{2} \Gamma}{\delta \varphi_{C}(z) \delta \varphi_{B}(y)}
 = - \Delta^{-1}_{BC}(y-z) + \Gamma_{BC}^{int}(y,z;\underline{\varphi})  
\ee
The $\bar{\Gamma}$ vertices can be computed in terms of the proper ones by
expanding (\ref{eqdigammabarra}) in field powers.

\subsection{Boundary conditions}
Following \cite{SU2} we assign different types of boundary conditions according to
the nature of couplings; since $\Pi\{\underline{\varphi};\Lambda\}$ contains 
both relevant
and irrelevant couplings, the former ones will be indicated by $\sigma(\Lambda)$ and
$\sigma'(\Lambda)$ whereas the latter ones by $\Sigma(\Lambda)$. The relevant
couplings having classical analogous, $\sigma(\Lambda)$, will be given their physical values at
$\Lambda=0$ and $\Lambda_{0} \rightarrow \infty$; the relevant couplings not
present in $S_{cl}$, $\sigma'(\Lambda)$, will be determined by requiring 
vertex functions to satisfy
the ST identities at the same physical point. Finally, the irrelevant
couplings, $\Sigma(\Lambda)$,
will be set to zero at the U.V. scale $\Lambda=\Lambda_{0}$.
In this way the fine-tuning problem is avoided.\\
As an example, we will examine the boundary conditions for the couplings
generated by introducing the fermion doublet in the action:
\be
\bea
{\ds \sigma_{m_{\Psi}}(\Lambda=0)=0 \quad \sigma_{\Psi}(\Lambda=0)=0 \quad
\sigma_{\bar{\Psi} A \Psi}(\Lambda=0)=g }\nonumber\\
{\ds \sigma'_{\bar{\lambda} c \Psi}(\Lambda=0)=\sigma'_{\bar{\Psi} c
\lambda}(\Lambda=0)=1 }
\eea
\ee
$\sigma'_{\bar{\lambda} c \Psi}(\Lambda=0)$ and $\sigma'_{\bar{\Psi} c
\lambda}(\Lambda=0)$ are computed by the following ST identity:
\be
\gamma_{\mu} \Gamma^{\bar{\Psi} A \Psi}(k,p,q) \Gamma^{k_{\mu} c}_{\mu}(p) -
\Gamma^{\bar{\Psi} \Psi}(k) \Gamma^{\bar{\lambda} c \Psi}(k,p,q) +
\Gamma^{\bar{\Psi} \Psi}(-q) \Gamma^{\bar{\Psi} c \lambda}(k,p,q) = 0
\footnote{The vertices must be computed at the point where the irrelevant parts
vanish.}
\ee
obtained by differentiating (\ref{relbrsgammaprimo}) with respect to the fields
$\bar{\Psi}$,$\Psi$ and $c$.

\section{Loop expansion: results}
It is possible to compute all the vertices of the theory, at least order by
order, by solving the RG equation iteratively.\\
One loop contributions can be derived by expanding the r.h.s. of
(\ref{equazionegruppo}) at the tree level:
\be \label{equazionegruppo0}
\ldl \Pi^{(1)}\{\underline{\varphi};\Lambda\} = \frac{1}{2} \int \frac{dq}{(2 \pi)^{4}} 
\Big[ \ldl D_{ED}(q;\Lambda) \Big] \bar{\Gamma}^{(0)}_{DE}(-q,q;\underline{\varphi})
\ee
Obviously, one loop corrections to the single vertex can be obtained by 
differentiating (\ref{equazionegruppo0}) with respect to the fields marking 
the vertex itself and, then, by solving the resulting equation.\\
We are going to examine in detail the fermionic sector, while the other results
will be only quoted. Then the $\beta$-function and the gauge symmetry recovery
will be taken into account.
\subsection{One loop couplings}
The corrections to the inverse fermion propagator take the form:
\be \label{fermione}
\Pi_{f}(p,\Lambda) = \sigma_{m_{\Psi}}(\Lambda) + \sigma_{\Psi}(\Lambda) (p 
\hspace{-0.52 em}/ + m) + \Sigma_{f}(p,\Lambda)
\ee
with $\Sigma_{f}(p \hspace{-0.52 em}/ = - m,\Lambda) = 0$ and 
$(\partial_{p_{\mu}} \Sigma_{f})(p=0,\Lambda) = 0$.
By substituting (\ref{fermione}) in (\ref{equazionegruppo0}) after 
differentiating it with respect to $\bar{\Psi} \Psi$ one easily obtains
\be
\bea
{\ds \sigma_{m_{\Psi}}(\Lambda) = \left. \int_{0}^{\Lambda} d\lambda 
\partial_{\lambda} \Pi_{f}(p,\lambda) \right|_{p \hspace{-0.48 em}/ = - m} }\nonumber\\
{\ds \gamma_{\mu} \sigma_{\Psi}(\Lambda) = \left. \partial_{p_{\mu}} \int_{0}^
{\Lambda} d\lambda \partial_{\lambda} \Pi_{f}(p,\lambda) \right|_{p = 0} }\nonumber\\
{\ds \Sigma_{f}(p,\Lambda = 0) = -\left[ \int_{0}^{\Lambda_{0}} d\lambda
\partial_{\lambda} \Pi_{f}(p,\lambda) - \sigma_{m_{\Psi}}(\Lambda_{0}) - 
\sigma_{\Psi}(\Lambda_{0}) (p \hspace{-0.52 em}/ + m) \right] }
\eea
\ee
In order to compute these integrals the Feynman parameter technique has been 
used; owing to some approximations concerning products of cutoff functions,
the following results hold only in the limit of large $\Lambda$.
\be
\bea
{\ds \sigma^{(1)}_{m_{\Psi}}(\Lambda) = \frac{3}{4} \left[ -\frac{3 g^{2}}{16 
\pi^{2}} m \log \frac{\Lambda^{2}}{m^{2}} \right] + {\cal O}(1) }\nonumber\\
{\ds \sigma^{(1)}_{\Psi}(\Lambda) = \frac{3}{4} \left[ -\frac{g^{2}}{16
\pi^{2}} \log \frac{\Lambda^{2}}{m^{2}} \right] + {\cal O}(1) }\nonumber\\
{\ds \Sigma^{(1)}_{f}(p,\Lambda = 0) = {\cal O}(1) }
\eea
\ee
The same technique can be used to compute all the other couplings\footnote{The 
form of the vertices involving these couplings is reported in (\ref{sint}).}:
\be
\bea
{\ds \sigma^{(1)}_{\omega c}(\Lambda) = \frac{g^{2}}{16 \pi^{2}} \log \frac{
\Lambda^{2}}{\mu^{2}} \qquad \Sigma^{(1)}_{\omega c}(\Lambda=0) = \frac{g^{2}}
{16 \pi^{2}} \log \frac{p^{2}}{\mu^{2}} - \frac{g^{2}}{16 \pi^{2}}
}\nonumber\\[0.5 cm]
{\ds \sigma_{m_{A}}^{(1)}(\Lambda) = \frac{- g^{2}}{16 \pi^{2}} \Lambda^{2} 
\quad \sigma_{\alpha}^{(1)}(\Lambda) = {\cal O}(1) \quad 
\sigma_{A}^{(1)}(\Lambda) =  \frac{g^{2}}{24 \pi^{2}} \left[ 5 \log
\left( \frac{\Lambda^{2}}{\mu^{2}} \right ) - \log
\left( \frac{\Lambda^{2}}{m^{2}} \right ) \right] }\nonumber\\[0.5 cm]
{\ds \lim_{m \rightarrow 0} \Sigma^{(1)}_{T} (p^{2}=\mu'^{2},\Lambda=0) = 
\frac{g^{2}}{6 \pi^{2}} \log \frac{\mu'^{2}}{\mu^{2}} \qquad \Sigma^{(1)}_{L}
(p, \Lambda=0) = {\cal O}(1) }\nonumber\\[0.5 cm]
{\ds \sigma^{(1)}_{\omega c A}(\Lambda) = -\frac{g^{2}}{16 \pi^{2}} \log \frac{
\Lambda^{2}}{\mu^{2}} \quad \left.\Sigma^{(1)}_{\omega c A}(l,p,k,\Lambda=0)
\right|_{3SP'} = \frac{g^{2}}{16 \pi^{2}} \log \frac{\mu^{2}}{\mu'^{2}} }
\nonumber\\[0.5 cm]
{\ds \sigma^{(1)}_{\bar{\Psi} A \Psi}(\Lambda) = \frac{g^{2}}{64 \pi^{2}}
\frac{11}{2} \log \frac{\Lambda^{2}}{m^{2}} \qquad \Sigma^{(1)}_{\bar{\Psi} A 
\Psi}(l,p,k,\Lambda=0) = {\cal O}(\frac{1}{\Lambda_{0}}) }\nonumber\\[0.5 cm]
{\ds \sigma'^{(1)}_{Lcc}(\Lambda) = \frac{g^{2}}{16 \pi^{2}} \log \frac{
\Lambda^{2}}{\mu^{2}} \qquad \Sigma'^{(1)}_{Lcc}((l,p,k,\Lambda=0) = {\cal O}(1) 
}\nonumber\\[0.5 cm]
{\ds \sigma'^{(1)}_{\bar{\lambda} c \Psi}(\Lambda) \equiv \sigma'^{(1)}_{\bar
{\Psi} c \lambda}(\Lambda) = -\frac{g^{2}}{16 \pi^{2}} \log \frac{\Lambda^{2}}
{\mu^{2}} \qquad 
\Sigma'^{(1)}_{\bar{\lambda} c \Psi}(\Lambda=0) \equiv \Sigma'^{(1)}_{\bar{
\Psi} c \lambda}(\Lambda=0) = {\cal O}(1) }
\eea
\ee
$\sigma^{(1)}_{3A}$ and $\sigma'^{(1)}_{3A}$ have not been computed since they
receive no contribution from radiative corrections involving matter fields. As
regards the couplings in the four vectorial field vertex, we only checked that
matter fields did not modify the $\Lambda$-power behavior.
\subsection{One loop ${\bf \beta}$-function}
Two renormalizable field theories, involving different subtraction points, can
be connected in such a way as to describe the same physics. Each one, indeed,
can be reconstructed from the other by introducing suitable counter terms. This
procedure is equivalent to redefining coupling constants and scale constants 
$Z$ so that the following scale equations are satisfied:
\be \label{relscala}
\Gamma_{nA,m \omega c,l \bar{\Psi} \Psi}(g',\mu') = Z_{A}^{-\frac{n}{2}} 
Z_{C}^{-m} Z_{\Psi}^{-l} \Gamma_{nA,m \omega c,l \bar{\Psi} \Psi}(g,\mu) 
\ee
where $\Gamma_{nA,m \omega c,l \bar{\Psi} \Psi}$ denotes the vertex with $n$ vector
fields, $m$ pairs $\omega_{\mu} c$ and $l$ pairs $\bar{\Psi} \Psi$ at the
physical point.\\
By using (\ref{relscala}) it is possible to write down three one loop equations,
involving $\Gamma_{AA}^{(0)}$, $\Gamma_{\Psi \bar{\Psi}}^{(0)}$ and 
$\Gamma_{\bar{\Psi} A \Psi}^{(0)}$, and then compute at the first
perturbative order $Z_{A}^{(1)}$ and $Z_{\Psi}^{(1)}$, as well the relation
between the new coupling constant $g'$ and the old one $g$.
\be \label{gprimo}
\bea
{\ds Z_{A}^{(1)} = 1 + \frac{g^{2}}{16 \pi^{2}} \frac{8}{3} \log \frac{\mu'^{2}}
{\mu^{2}} \qquad Z_{\Psi}^{(1)} = \frac{g'}{g} \left[ 1 + \frac{g^{2}}
{16 \pi^{2}} \log \frac{\mu'^{2}}{\mu^{2}} \right] }\nonumber\\[0.6 cm]
{\ds g' = \frac{g - \frac{g^{2}}{16 \pi^{2}} \log \frac{\mu'^{2}}{\mu^{2}} }
{\sqrt{Z_{A}^{(1)}} \left( 1 + \frac{g^{2}}{16 \pi^{2}} \log \frac{\mu'^{2}}
{\mu^{2}} \right) } }
\eea
\ee
By differentiating (\ref{gprimo}), the $\beta$-function in the zero mass limit
\footnote{We neglect input mass parameter in the limit of large momenta.} is 
obtained
\be
\beta^{(1)}(g) \equiv \mu' \left. \frac{\partial g'}{\partial \mu'} \right|_{\mu'=
\mu} = -\frac{5}{12 \pi^{2}} g^{3}
\ee
The same result can be computed by using other scale equations, involving 
different vertices.
\subsection{Gluon transversality}
We will show a check of the ST identity involving the inverse vector and ghost 
propagators at one loop order. Differentiating (\ref{relbrsgammaprimo}) with
respect to the fields $A_{\mu} c$ gives rise to
\be
\left[ \Gamma_{\mu \nu}(p, \Lambda=0) + p_{\mu} p_{\nu} \right] \Gamma_{\mu}^{
k_{\mu} c} (p, \Lambda=0) = 0
\ee
which forces the longitudinal irrelevant part of the inverse vector propagator,
$\Sigma_{L}(p, \Lambda)$, to be zero at the physical point.\\
One finds
\be
\lim_{\Lambda_{0} \rightarrow \infty} \Sigma_{L}^{(1)}(p, \Lambda=0) = -
\int^{\infty}_{0} d\lambda \partial_{\lambda} \left[ \Pi_{L}(p,\lambda) - 
\Pi_{L}(0,\lambda) - p^{2} \left. \frac{\partial}{
\partial \bar{p}^{2}} \Pi_{L}(\bar{p},\lambda) \right|_
{\bar{p}^{2} = \mu^{2}} \right] = 0
\ee
Owing to the irrelevant nature of $\Sigma_{L}$, the leading term as $\Lambda_
{0} \rightarrow \infty$ can be nothing but a constant; so it is 
sufficient to show this constant to be zero.
\subsection{Perturbative renormalizability}
In order to prove the theory to be perturbatively renormalizable one has to 
show that RG equation has a nontrivial limit as $\Lambda_{0} \rightarrow 
\infty$.\\
Starting from the integral form of the RG equation (\ref{equazionegruppo}), so 
as to embody boundary conditions, we analyse the behavior at large scales of
proper vertices, whose form can be obtained, order by order, by successive 
iterations. If their $\Lambda$-dependence is such that power counting is 
satisfied at any order, perturbative renormalizability is proved.\\
In order to simplify dimensional analysis by neglecting every momentum 
dependence, we introduce the norm \cite{Po}:
\be
\left| \Gamma(n_{A}, n_{\bar{c} c}, n_{\bar{\Psi} \Psi}, n_{k_{\mu}}, n_{L},
n_{\bar{\lambda} \lambda}) \right|_{\Lambda} \doteq \max_{p_{i}^{2} \leq c
\Lambda^{2}} \left| \Gamma(p_{1},\ldots,p_{n};\Lambda) \right|
\ee
where the generic $n_{\varphi}$ is the number of $\varphi$ fields or sources
marking the vertex.\\
Since 
\be
dim \Gamma_{n_{\varphi_{1}} \cdots n_{\varphi_{n}}}(p_{1},\ldots,
p_{n}) = 4 - n_{A} - 
2n_{\bar{c}c} - 3n_{\bar{\Psi} \Psi} - 2n_{k_{\mu}} - 2n_{L}- 3n_{\bar{\lambda}
\lambda} \doteq 4 - \sum_{i} \alpha_{i} n_{\varphi_{i}}
\ee
 we assume, at $l$-loop order and for large $\Lambda$, that
\be \label{assun}
\left| \Gamma^{(l)}( n_{\varphi_{1}},\cdots, n_{\varphi_{n}}) \right|_{\Lambda} = {
\cal O}(\Lambda^{4 - \sum_{i} \alpha_{i} n_{\varphi_{i}} })
\ee
We want to show this $\Lambda$-dependence to be preserved at $(l+1)$-loop 
order.\\
Since, by definition, auxiliary vertices have the same dimension as the proper
ones, the following relations hold:
\be \label{relazioni}
\left| \bar{\Gamma}^{(l)}_{a b n_{\varphi_{1}} \cdots n_{\varphi_{n}}}
\right|_{\Lambda} = \left\{ \bea
{\cal O}(\Lambda^{4 - \sum_{i} \alpha_{i} n_{\varphi_{i}} - 2})  \\[0.7 cm]
{\cal O}(\Lambda^{4 - \sum_{i} \alpha_{i} n_{\varphi_{i}} - 3})
\eea
\right.
\ee
where the upper relation holds if $ab$ refer to ghost or vectorial fields, while
the lower holds if $ab$ refer to matter fields.\\
From (\ref{equazionegruppo}) and (\ref{relazioni}) one gets
\be
\left| \Pi^{(l + 1)}_{n_{\varphi_{1}} \cdots n_{\varphi_{n}}} \right|_
{\Lambda} = \left\{ \bea
{\cal O}(\Lambda^{2 - \sum_{i} \alpha_{i} n_{\varphi_{i}} + 4 - 2})\\[0.7 cm]
{\cal O}(\Lambda^{1 - \sum_{i} \alpha_{i} n_{\varphi_{i}} + 4 - 1})
\eea
\right.
\ee
by taking into account the different contributions $M_{ED}$ gives according to
the nature of the $ED$ fields as well as the different integration domains, 
related to the relevant and irrelevant parts $\Pi$ is made by (see also
\cite{QED} for QED).\\
Since (\ref{assun}) is satisfied at $l=0$, by induction on the loop number we
prove perturbative renormalizability. 
\section{Conclusions}
A RG application to a non-abelian gauge theory interacting with a massive
doublet by a minimal coupling has been shown.\\
We analysed in detail a mathematical technique which enables the Green function
determination when the regularization breaks the gauge symmetry. This approach
is, therefore, very useful to deal with chiral gauge theories, since in this
case symmetry breaking, due to the regularization, is unavoidable.
Moreover this method let us obtain an exact RG equation, describing the
effective action evolution at any energy scale. Already in 1983 Becchi \cite{Be}
proposed an exact RG equation in differential form together with boundary
conditions fixed on the bare couplings: this fixing gave rise to the so called
fine-tuning problem concerning the determination of the boundary conditions
related to couplings not present in the classical action. In order to avoid
this problem, boundary conditions are assigned on the renormalized theory.\\
By using this technique we computed, at one loop order, all the couplings at
large scales: they perfectly agree with the results obtained by dimensional
regularization. By evaluating couplings at the U.V. scale one achieves the
relations between bare and renormalized couplings.\\
The gauge symmetry recovery at the physical point has been analysed: we
checked, at one loop order, gluon transversality as a consequence of one of the ST 
identities; it would be interesting to verify the ST identities to be satisfied at
any loop.\\
Finally we computed the one loop $\beta$-function, describing the asymptotic
behavior of the theory. This result coincides with the one obtained by
dimensional regularization in the independent mass prescription.\\
 
\section*{Acknowledgments}
We wish to thank M. Bonini, F. Guerra and M. Testa for their willingness to helpful
discussions.

\end{document}